\documentclass[preprint,aps,pra,showpacs]{revtex4}
\usepackage{tabularx}
\usepackage{dcolumn}
\usepackage{graphics}
\usepackage{bm}
\usepackage[dvips]{graphicx}
\usepackage[dvips]{rotating}
\usepackage[latin1]{inputenc}
\usepackage{dcolumn}
\usepackage{tabularx}
\usepackage{amsmath}
\usepackage{amssymb}
\begin{document}
\draft

\title{Oscillations in the total photodetachment cross sections of a triatomic anion}

\author{B.C. Yang and  M.L. Du} \email{
duml@itp.ac.cn} \affiliation{Institute of Theoretical Physics,
Academia Sinica, Beijing 100190, China}



\date{\today}

\begin{abstract}

The total photodetachment cross section of a linear triatomic
anion is derived for arbitrary laser polarization direction.
The cross section is shown to be strongly oscillatory when the laser
polarization direction is parallel to the axis of the system; the oscillation amplitude decreases and vanishes as the angle between the laser polarization and the anion axis increases and becomes perpendicular to the axis. The average cross section over the orientations of the triatomic system is also obtained. The cross
section of the triatomic anion is compared with the cross
section of a two-center system. We find there are two oscillation frequencies in the triatomic anion in contrast to only one oscillation frequency in the two-center case. Closed-orbit theory is used to
explain the oscillations.

\par


\end{abstract}

\maketitle


\section{Introduction}

Photodetachment of negative ions in the presence of a static
electric field has been an active research area in the last decades
\cite{Bryant,Stewart,Du1,Du2,Du3,Gibson1,Gibson2,Rangan,Gibson3,Fabrikant1,Rau1,Fabrikant2,
Fabrikant3,Rau2,Afaq1,Kramer,Brache1,Bracher2,Gao, Blondel1,Wang,
Fabrikant4,Blondel2,Du7,Du8}. The most interesting feature in the
cross section is the induced oscillations above photodetachment
threshold by the static electric field. The oscillations in the
total cross sections can be understood using closed-orbit theory
\cite{Du4, Du5, Du6}. In an effort to understand the oscillations in
various processes involving two-center system such as the
photoionization cross section of diatomic molecule \cite{Fano1966},
the scattering of $D_2$ molecule by fast electron
\cite{Kamalou2005}, a molecule in a strong laser field
\cite{Hu2005}, above-threshold ionization \cite{Xi2008} and harmonic
generation \cite{Hetz2007}, Afaq and Du extended the one-center H$^-$ model for photodetachment and
developed a two-center model for photodetachment\cite{Afaq2, Afaq3,
Du09}. They demonstrated the cross sections in the two-center system
show strong oscillation which can be explained using closed-orbit
theory. In particular, a detached electron orbit connecting the two
centers is identified to be responsible for the oscillation in the
total photodetachment cross sections of the two-center system.

To understand the structural information on linear triatomic
negative anions such as BeCl$^-_2$,HCN$^-$,CS$^-_2$ and
CO$^-_2$\cite{Rajgara}, Afaq \emph{et al.}\cite{Afaq4} recently
studied the photodetachment of a triatomic anion with three centers
when the axis of the triatomic ion is perpendicular to the laser
polarization direction. Interference patterns for detached-electron
on a screen placed at a large distance from the system were
demonstrated, but the total cross section was found to be smooth and
no oscillation was observed for this configuration.

Here we extend the study of the total photodetachment cross section
of the above triatomic anion system to the general case with an
arbitrary laser polarization direction. We will derive analytic
formulas for the total cross section which depends on photon energy,
laser polarization direction and other parameters characterizing the
triatomic anion. It will be shown that the cross section shows
strong oscillations when the laser polarization is parallel to the
system axis and the oscillation amplitudes gradually decrease to
zero as the laser polarization is changed to be perpendicular to the
axis. We also obtained the cross section averaged over the
orientations of the system. We compare the cross section of the
triatomic anion with that of the two-center system. The oscillations
in photodetachment cross sections for the  triatomic anion appear
much enhanced compared to the two-center case. We find there are two
oscillation frequencies in the triatomic anion. The two oscillations
are explained using closed-orbit theory. Atomic units will be used
unless specially noted.

\section{Formulas for total photodetachment cross section}

The linear triatomic anion interacting with a laser is shown
schematically in Fig.1. Symbols 1,0 and 2 represent the three atomic
centers in the system. It is convenient to choose the z-axis in the
direction of the three-center axis and the middle center denoted by
0 as the origin of coordinates. Let $d$ be the distance between two
adjacent centers. The laser polarization direction is denoted as
$(\theta_L,\phi_L)$ with respect to the $z$ axis.

In the triatomic anion, one active electron is assumed. This is an
extension of the two-center model   for photodetachment\cite{Afaq2,
Afaq3, Du09} to a three-center model by Afaq \emph{et
al.}\cite{Afaq4}. In the photodetachment process, there are two
steps: in the first step, the active electron absorbs one photon
energy $E_{ph}$ and escapes from the negative anion as an
electron wave from each center; in the second step, the outgoing
waves from each center propagate out to large distances. The
interference of the outgoing waves from each center produces
oscillatory cross section.

For the general case, as illustrated in Fig. 1, Let $\Psi^+_{1}$,
$\Psi^+_{0}$ and $\Psi^+_{2}$ be the detached-waves from center 1, 0
and 2 respectively. Following the previous approach in the
two-center case\cite{Du09}, the outgoing detached-electron wave
$\Psi^{+}_M$ from the triatomic anion can be written as a linear
combination given by\cite{Du09}
\begin{equation}
\Psi^{+}_M=\frac{1}{\sqrt{3}}(\Psi^+_{1}+\Psi^+_{0}+\Psi^+_{2}).
\end{equation}
Let $(r_{1},\theta_{1},\phi_{1})$, $(r_{0},\theta_{0},\phi_{0})$ and
$(r_{2},\theta_{2},\phi_{2})$ represent the spherical coordinates of
the detached-electron relative to the three centers respectively.
The detached-electron wave generated from each center has been worked previously\cite{Du8}. They can be written as
\begin{eqnarray}
  \Psi_1^{+}&=&\frac{4Bk^2i}{(k_b^2+k^2)^2}f(\theta_{1},\phi_{1};\theta_{L},\phi_{L})\frac{\exp(ikr_1)}{kr_1}, \nonumber \\
  \Psi_0^{+}&=&\frac{4Bk^2i}{(k_b^2+k^2)^2}f(\theta_{0},\phi_{0};\theta_{L},\phi_{L})\frac{\exp(i kr_0)}{kr_0}, \nonumber \\
  \Psi_2^{+}&=&\frac{4Bk^2i}{(k_b^2+k^2)^2}f(\theta_{2},\phi_{2};\theta_{L},\phi_{L})\frac{\exp(ikr_2)}{kr_2},
\end{eqnarray}
where $k=\sqrt{2E}$ and E is the detached-electron energy; $k_b$ is
related to the binding energy $E_b$ of H$^-$ by
$E_b=\frac{k_{b}^2}{2}$. The photon energy is given by
$E_{ph}=E+E_b$. $B$ is a normalization constant\cite{Du1}, $c$ is
the speed of light and is approximately equal to 137a.u.. The
angular factor such as $f(\theta_{0},\phi_{0};\theta_{L},\phi_{L})$
represents the dependence of the detached-electron wave function on
the outgoing direction $(\theta_{0},\phi_{0})$ for center 0, and it
can be written as \cite{Du8}
\begin{equation}\label{1}
f(\theta_{0},\phi_{0};\theta_{L},\phi_{L})=\cos\theta_0\cos(\theta_L)+\sin\theta_0\sin(\theta_L)\cos(\phi_0-\phi_L).
\end{equation}
The angular factors $f(\theta_{1},\phi_{1};\theta_{L},\phi_{L})$
and $f(\theta_{2},\phi_{2};\theta_{L},\phi_{L})$ can be written out
in a similar way.

After substituting Eqs.(2) and Eqs.(3) in Eq.(1), we can get the explicit expression for the detached-electron wave
function $\Psi^{+}_M$. The resulting formula for $\Psi^{+}_M$ can be simplified because the calculation of the
photodetachment cross section requires the knowledge of $\Psi^{+}_M$
at large distances where $r_1$, $r_0$ and $r_2$ are much
greater than the distance $d$ between two neighboring centers.
Let $(r, \theta, \phi)$
be the spherical coordinates of the detached-electron relative to
the origin. Then we approximate the phase terms using $r_1\approx
r-d\cos\theta$, $r_0=r$, $r_2\approx r+d\cos\theta$; in other places
we can set $r_1\approx r_2\approx r_0=r$,
$\theta_1\approx\theta_2\approx\theta_0=\theta$. With these
approximations, $\Psi^{+}_M$ becomes
\begin{equation}\label{3}
    \Psi_M^+(r, \theta,
    \phi)=\frac{4Bk^2i}{(k_b^2+k^2)^2}\frac{1}{\sqrt{3}}
    f(\theta,\phi;\theta_{L},\phi_{L})[1+2\cos(kd\cos\theta)]
    \frac{\exp(ikr)}{kr},r\rightarrow\infty.
\end{equation}
Eq.(4) describes the detached-electron wave of the linear triatomic anion when the detached-electron is far away from the anion.

The differential cross section is \cite{Du3}
\begin{eqnarray}
       \frac{d\sigma(\textbf{q})}{ds}=\frac{2\pi E_{ph}}{c}\textbf{j}\cdot\textbf{n},\\
       \nonumber \textbf{j}=\frac{i}{2}(\Psi_M^+\nabla\Psi_M^{+\ast}-\Psi_M^{+\ast}\nabla\Psi_M^+),
\end{eqnarray}
where $\sigma$ is the photodetachment cross section, $ds$ is the
differential area on a surface $\Gamma$ such as the surface of a
large sphere enclosing the anion, $\textbf{q}$ is the
coordinate on the surface $\Gamma$, $\textbf{n}$ is the exterior
norm vector at $\textbf{q}$ and $\textbf{j}$ is the detached-electron
flux.

The detached-electron flux in the radial direction can be evaluated
as
\begin{equation}\label{5}
    j_r(r,\theta, \phi)=\textbf{j}\cdot\hat{\textbf{r}}=\frac{i}{2}(\Psi_M^+\frac{\partial\Psi_M^{+\ast}}{\partial r}-\Psi_M^{+\ast}\frac{\partial\Psi_M^+}{\partial r}).
\end{equation}
When Eq.(4) is substituted in Eq.(6), we get
\begin{equation}\label{6}
    \textbf{j}\cdot\hat{\textbf{r}}=\frac{16B^2k^4}{3k(k_b^2+k^2)^4}\frac{f^2(\theta,\phi;\theta_{L},\phi_{L})}{r^2}[1+4\cos(kd\cos\theta)+4\cos^2(kd\cos\theta)].
\end{equation}

To get the total photodetachment cross section, we integrate
the differential photodetachment cross section using Eq.(5) and Eq.(7)  over $\theta$ and $\phi$,
\begin{equation}
    \sigma(E, d, \theta_L)=\frac{2\pi E_{ph}}{c}
    \int(\textbf{j}\cdot\hat{\textbf{r}})r^2\sin\theta d\theta
    d\phi.
\end{equation}
After a straightforward integration, we find the total
photodetachment cross section can be written as a product form
\begin{equation}\label{7}
    \sigma(E, d, \theta_L)=\sigma_0(E)A_3(kd, \theta_L),
\end{equation}
where
\begin{eqnarray}
  \sigma_0(E)&=&\frac{16\sqrt{2}B^2\pi^2E^\frac{3}{2}}{3c(E_b+E)^3}, \\
  A_3(kd, \theta_L)&=&1+\frac{4}{3}I(kd)+\frac{2}{3}I(2kd).
\end{eqnarray}
In the above equations, $\sigma_0(E)$ is the smooth total
photodetachment cross section of H$^-$\cite{Du1} and $A_3(kd,
\theta_L)$ is a modulation function for the triatomic anion. The
function $I(S)$ appearing in Eq.(11) is given by
\begin{equation}
I(S)=3\cos^2\theta_L[\frac{\sin S}{S}+3\frac{\cos
S}{S^2}-3\frac{\sin S}{S^3}]+3[\frac{\sin S}{S^3}-\frac{\cos
S}{S^2}].
\end{equation}
The function $I(S)$ also appears in the modulation function for the
two-center problem studied previously\cite{Du09}.

\section{Oscillations and limits of cross sections}

First we show the cross section obtained by Afaq $et$
$al$.\cite{Afaq4} for the perpendicular configuration can be
obtained from the general formulas. When the laser polarization
direction is perpendicular to the direction of the axis,
$\theta_L=\frac{\pi}{2}$. From Eqs.(9)-(12) we immediately have
\begin{equation}
    \sigma(E, d,\frac{\pi}{2})=\sigma_0(E)
    [1+4[\frac{\sin(kd)}{(kd)^3}-\frac{\cos(kd)}{(kd)^2}]
    +4[\frac{\sin(2kd)}{(2kd)^3}-\frac{\cos(2kd)}{(2kd)^2}]].
\end{equation}
The result in Eq.(13) is identical to that given by Afaq \emph{et
al.}\cite{Afaq4}. In Fig.2 we compare the photodetachment cross
section in Eq.(13) with the cases $\theta_L=0$ and
$\theta_L=\frac{\pi}{4}$ in Eqs.(9)-(12) when $d=4a_0$. Indeed we do
not find any oscillation in the the cross section for
$\theta_L=\frac{\pi}{2}$ in agreement with Afaq \emph{et
al.}\cite{Afaq4}. But when the direction of laser polarization is
not perpendicular to the axis, we observe a hint of oscillation as
illustrated in the insert. In fact, the oscillations will become
strong and obvious as the parameter $S=kd$ is increased to be
comparable or greater than $\pi$. In Fig.3 we show the
photodetachment cross sections in Eqs.(9)-(12) for several $d$ and
$\theta_L$ values. The following points can be derived directly from
Fig.3. First, we do not observe obvious oscillation in the
photodetachment cross sections for the perpendicular laser
polarization ($\theta_L=\frac{\pi}{2}$) even when $d$ is increased.
But when the laser polarization direction is not perpendicular to
the anion axis, there are strong oscillations. Second, the
oscillation amplitude depends on the laser polarization direction.
In fact, the oscillation amplitude increases and reaches maximum as
the laser polarization direction is turned from perpendicular to
parallel direction with respect to the anion axis ($\theta_L=0$).
However, the oscillation frequency is not sensitive to the laser
polarization direction. The modulation function $A_3$ suggests the
oscillation frequency increases but the oscillation amplitude
decreases as the parameter $d$ is increased. The oscillations are
easily observed when $d$ is approximately around $30-100a_0$ in our
system.

The limits of the photodetachment cross sections of the anion can be
obtained from the limits of $I(S)$. Using Taylor series
expansion,one can show
\begin{eqnarray}
    \lim_{S\rightarrow0}I(S,\theta_L)&=&1,\nonumber\\
     \lim_{S\rightarrow\infty} I(S,\theta_L)&=&0.
\end{eqnarray}
The above results are independent of the value of $\theta_L$.
Therefore we conclude from Eqs.(9)-(12) that in the low energy limit the total
photodetachment cross section of the three-center system is three
times of the photodetachment cross section of a single-center
system and in the large photon energy limit the photodetachment cross
section of the three-center system approaches the photodetachment
cross section of a single-center system.

It is also straightforward to get the photodetachment cross section
averaged over the orientations of the anion. Let us assume the
direction of the anion is random with respect to the laser
polarization direction. Defining the averaging by
\begin{equation}
    \bar{\sigma}(E, d)=\int_0^\pi\sin\theta_Ld\theta_L\int_0^{2\pi}\sigma(E, d, \theta_L)d\phi_L/\int_0^\pi\sin\theta_Ld\theta_L\int_0^{2\pi}d\phi_L.
\end{equation}
The integrals can be evaluated to give the following results,
\begin{eqnarray}
  \bar{\sigma}(E, d)=\sigma_0(E)\bar A_3(kd), \\
  \bar A_3(kd)=1+\frac{4}{3} \frac{\sin(kd)}{kd}+\frac{2}{3} \frac{\sin(2kd)}{2kd}.
\end{eqnarray}

It is interesting that the average cross section is equal to the
laser polarization dependent cross section in Eqs.(9)-(12) evaluated
at a special angle satisfying $\theta_L=\arccos(\frac{1}{\sqrt{3}})$
(approximately $54.74^\circ$). The same angle was noticed for the
two-center system earlier\cite{Du09}.

\section{Fourier analysis}

The averaging process does not change the basic oscillatory
structure of the cross section. However, the oscillation amplitudes
in the average cross section are reduced to one third of the
values obtained for the laser polarization dependent cross section
at $\theta_L=0$. In Fig.4(a) we compare the average cross section
and the laser polarization dependent cross section at $\theta_L=0$.

The oscillations are better analyzed using a transformation. For the cross section $\sigma(E)$ we define $F(x)$ using the following integral
\begin{equation}
    F(x)=\int_{E_1}^{E_2}[\sigma(E)-\sigma_0(E)]
    \sin[\frac{\pi(E-E_1)}{(E_2-E_1)}]k\exp(-ikx)dE.
\end{equation}

In Fig.4(b) we present the corresponding transformations of the two
cross sections in Fig.4(a). Several points can be derived from Fig.4
regarding the oscillations. First, there are two peaks in each
transformation. The  peaks correspond to two oscillations in each
cross section. Second, the oscillation frequencies are not changed
by the averaging procedure. Third,  the oscillation amplitudes in
the average cross section are considerable reduced.

We now compare the average photodetachment cross section of the
triatomic anion with that of the two-center negative ion studied
earlier\cite{Du09}. In Fig.5(a) we show the two cross sections. One
can see that the oscillation in the photodetachment cross section
for the triatomic anion is enhanced compared to that for the
two-center system. In Fig.5(b) we show the corresponding
transformations of the two cross sections in Fig.5(a). It is clear
from Fig.5(b) that there are two oscillations for the triatomic
anion but only one oscillation for the two-center system.

Closed-orbit theory was extended previously to explain the
oscillation in the cross section of the two-center system. The
oscillation  was identified as an interference between the
detached-electron wave emitted at one center and the source of the
wave at another center\cite{Du09}. A similar explanation can be made
here. For example, the detached-electron wave produced at center 1
will reach center 0 and 2 as it propagates out. The overlap of this
detached-electron wave from center 1 with the source at center 0 and
the source at 2 produce the $\sin(kd)$ and $\sin(2kd)$ oscillations
in the total cross section. Detached-electron wave from center 2
also have similar interference terms. The final oscillation
amplitudes include all the interference terms. We will not repeat
the detail of such derivation which is quite similar to the
two-center case\cite{Du09}. We emphasize that such a derivation
based on closed-orbit theory establishes that $kd$ is the action
$\int\textbf{p}\cdot d\textbf{q}$ of the detached-electron
propagating from one center to a neighboring center and $2kd$ is the
action of the detached-electron propagating from center 1 to center
2 or from center 2 to center 1. The two oscillations are directly
associated with the detached-electron orbits from one center to a
neighboring center having action $kd$ and the detached-electron
orbits connecting center 1 and 2 having action $2kd$.

\section{Conclusions}

We have studied the photodetachment of the recent triatomic anion
model\cite{Afaq4} in the general case. We have derived the total
photodetachment cross section for arbitrary laser polarization
direction. It is demonstrated there are two oscillation frequencies
in the cross sections. The amplitudes of the oscillations can be
varied by changing the laser polarization direction. The amplitudes
are largest when the laser polarization is parallel to the anion
axis. As the laser polarization direction is turned to be
perpendicular to the axis, the oscillation amplitudes decrease and
vanish. We also obtained the average cross section for random
orientations of anion. The averaging procedure modifies the
oscillation amplitudes but it does not change the oscillation
frequencies. The two oscillations in the present three-center system
can be explained using closed-orbit theory as interference between
detached-electron waves produced from one center and the sources at
other centers. Two types of detached-electron orbits are responsible
for the two oscillations.


\newpage
\begin{figure}
\includegraphics[scale=.80,angle=-0]{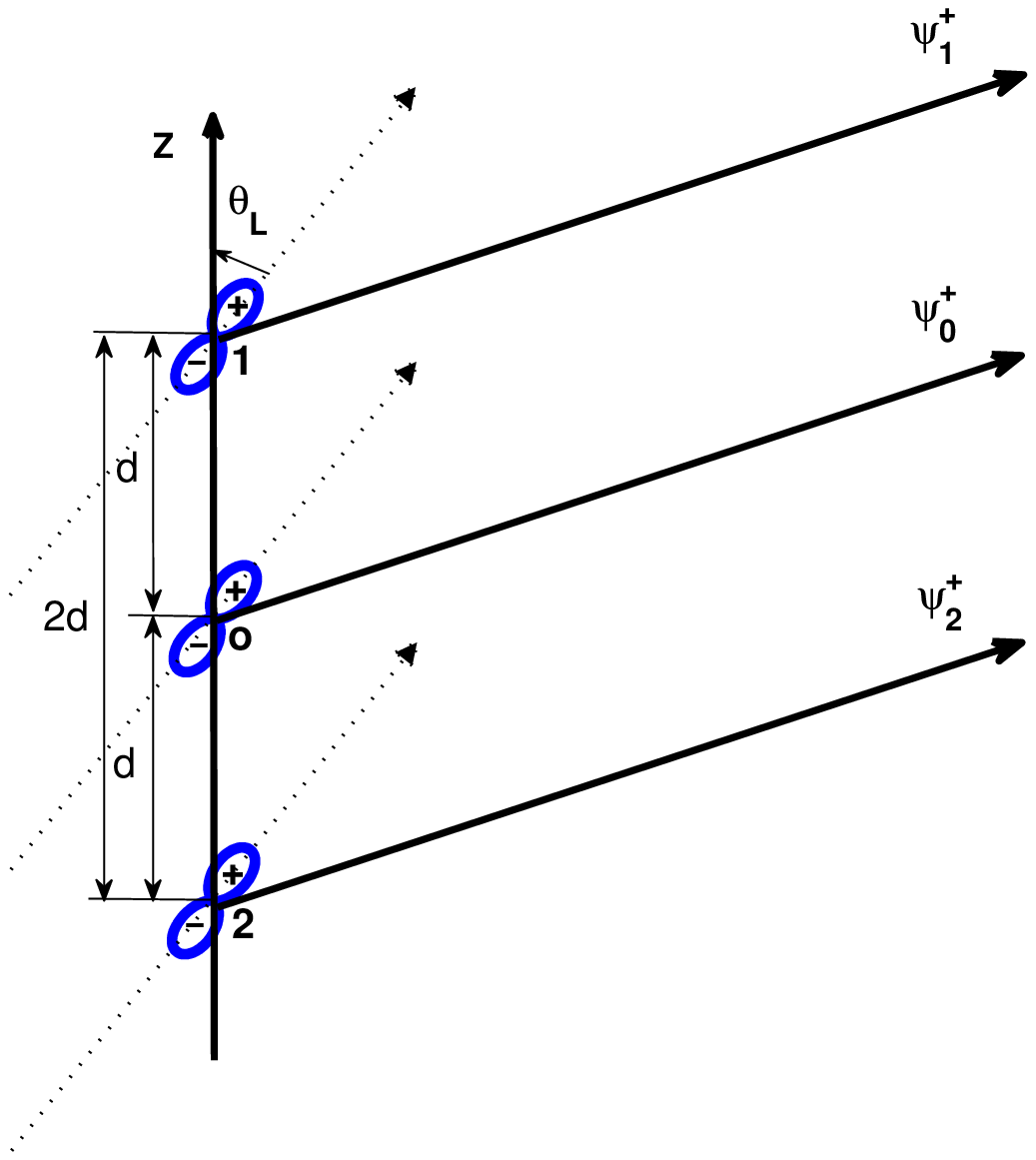}
\caption{The schematic representation for the photodetachment of the linear triatomic anion. The dotted
arrows point to the laser polarization direction. The plus and minus lobes represent the angular amplitude of the detached-electron from each center.}
\end{figure}

\begin{figure}
\includegraphics[scale=.80,angle=-0]{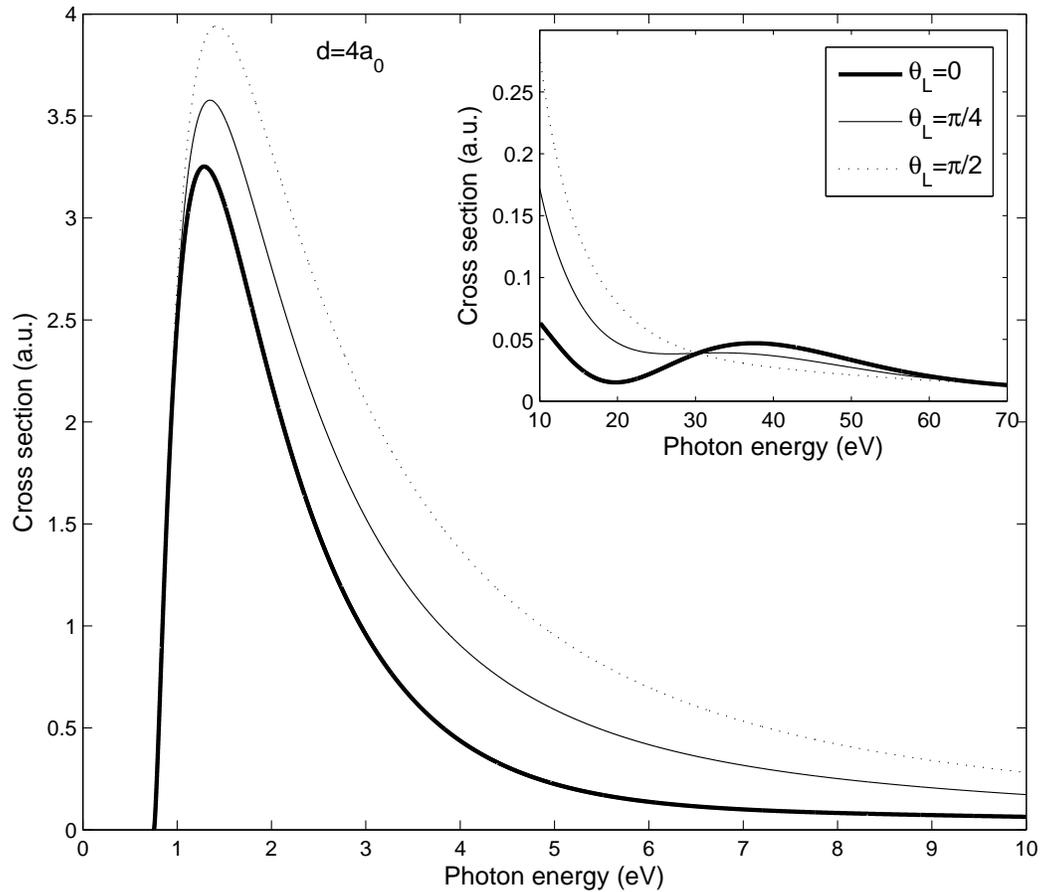}
\caption{The photodetachment cross section of the triatomic anion at
$d=4a_0$ and the laser polarization direction $\theta_L$ are
respectively  equal to $0$, $\frac{\pi}{4}$, $\frac{\pi}{2}$. The
inset provides a hint of oscillation for $\theta_L=0$ and
$\theta_L=\frac{\pi}{4}$.}
\end{figure}

\begin{figure}
\includegraphics[scale=.80,angle=-0]{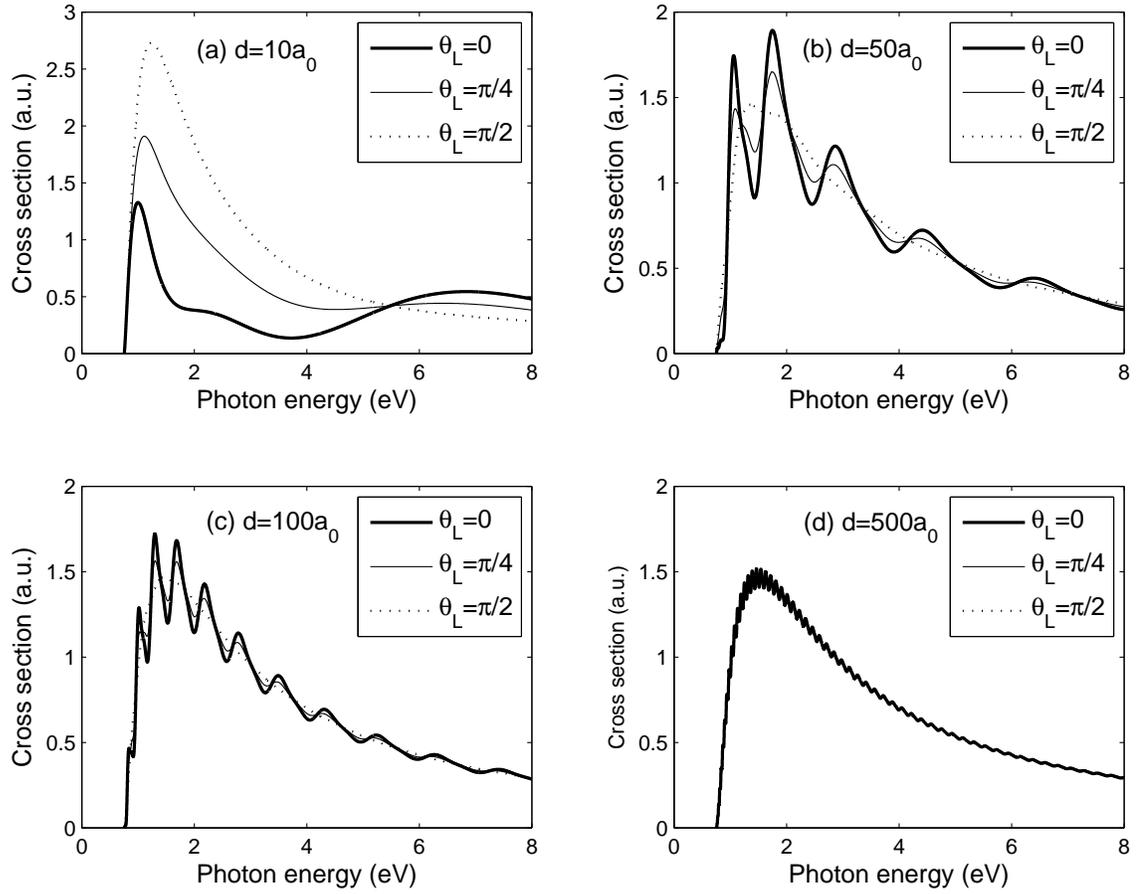}
\caption{The photodetachment cross sections of the triatomic anion with
different values of $d$ and $\theta_L$. }
\end{figure}

\begin{figure}
  \includegraphics[scale=.80,angle=-0]{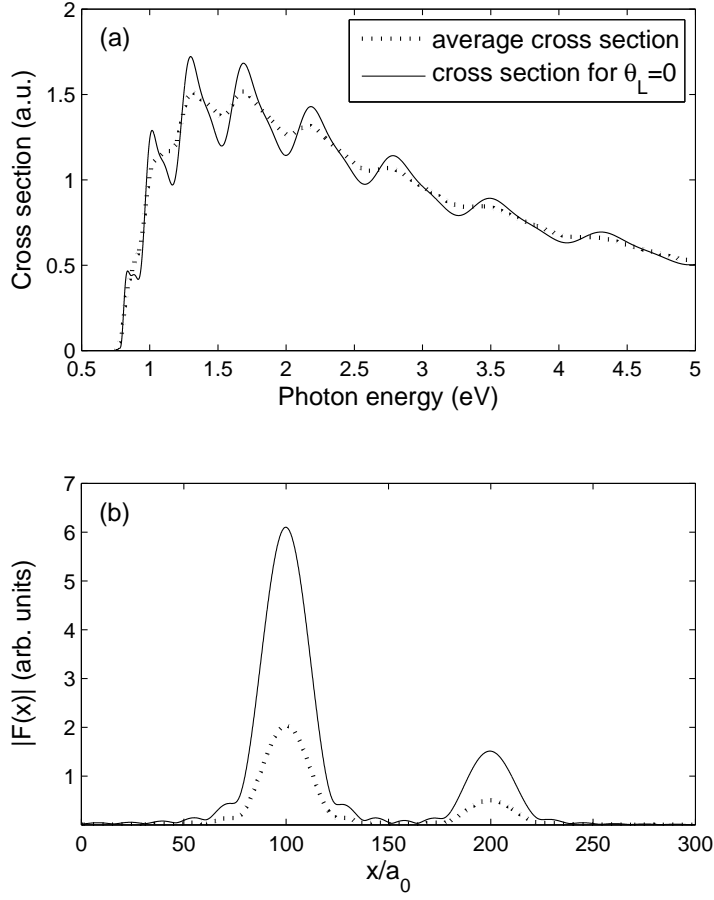}
  \caption{(a) The total photodetachment cross section for $d=100a_0$ and $\theta_L=0$ (solid line) and the average cross section (dotted line).   (b) The transformations defined in Eq.(18) for the above two cross sections.}
\end{figure}

\begin{figure}
\includegraphics[scale=.80,angle=-0]{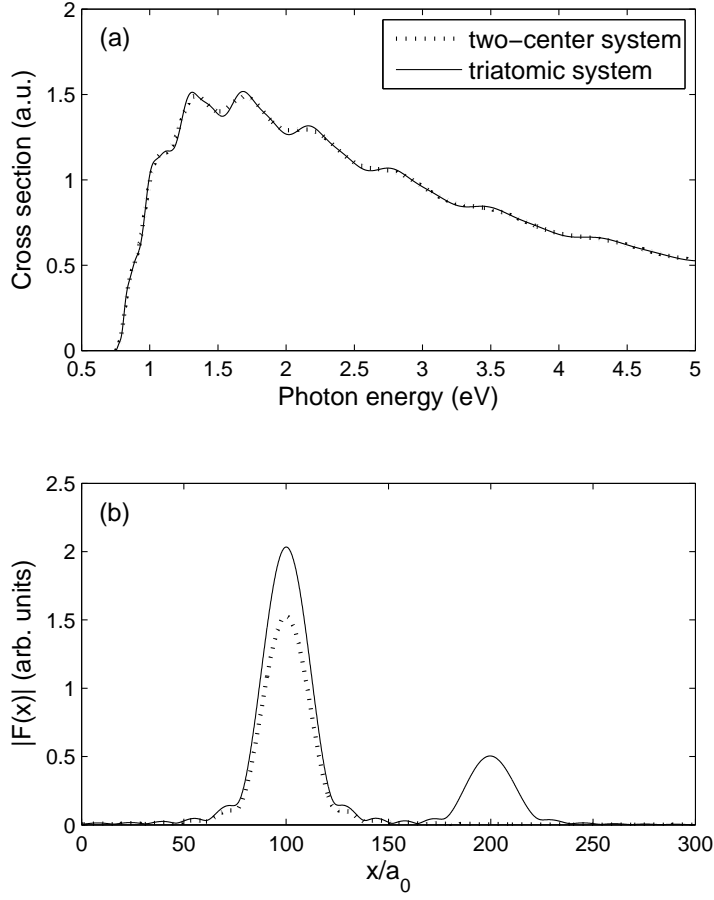}
\caption{(a) The orientation average photodetachment cross
sections of the two-center system and the orientation average triatomic anion system.
The parameter $d=100a_0$. (b) The transformations defined in Eq.(18) for  the above
two cross sections.}
\end{figure}


\end{document}